\documentclass[prb,twocolumn]{revtex4}

\usepackage{graphicx}
\begin{document}

\title{Room Temperature Domain Wall Pinning in Bent Ferromagnetic
Nanowires}
\date{\today}
\author{D. M. Silevitch}
\author{M. Tanase}
\author{C. L. Chien}
\author{D. H. Reich}
\email[Contact:]{dhr@pha.jhu.edu}
\affiliation{Department of Physics and
Astronomy, The Johns Hopkins University, Baltimore, MD 21218}

\begin{abstract} 
Mechanically bent nickel nanowires show clear features in their room 
temperature magnetoresistance when a domain wall is pinned at the location 
of the bend. By varying the direction of an applied magnetic field, the wire 
can be prepared either in a single-domain state or a two-domain state. 
The presence or absence of the domain wall acts to shift the switching 
fields of the nanowire. In addition, a comparison of the magnetoresistance 
of the nanowire with and without a domain wall shows a shift in the 
resistance correlated with the presence of a wall. The resistance is 
decreased by $20-30\:\mathrm{m}\Omega$ when a wall is present, 
compared to an overall resistance of $40-60\:\Omega$. A model of the 
magnetization was developed that allowed calculation of the magnetostatic 
energy of the nanowires, giving an estimate for the nucleation energy of a domain wall.
\end{abstract}

\maketitle 

\section{Introduction} 

Electrodeposited nanowires provide  useful systems for studying electrical 
and magnetic phenomena at sub-micron length scales. Metallic nanowires with 
diameters in the 20--500 nm range can be grown in bulk quantities to lengths 
as long as 50 microns.\cite{Fert99} Since the composition of the nanowires can be 
varied along their length, they can be used to study phenomena where electrical 
current flows perpendicular to the composition modulation. There has been 
significant interest in studying the magnetic properties of individual pure 
ferromagnetic nanowires, including the mechanism and dynamics of magnetization 
reversal,\cite{Wernsdorfer96} spin transfer,\cite{Jaccard00} and domain wall 
propagation.\cite{Radulescu00} A variety of techniques have been used to examine 
nanowires, including micro-SQUID,\cite{Wernsdorfer96} magnetic force
microscopy,\cite{Lederman95,Radulescu00} and 
magnetoresistance\cite{Radulescu00,Wegrowe99,Jaccard00} techniques. 
Previous magnetoresistance measurements have primarily focused on wires 
that are still embedded in their fabrication templates, using special growth 
techniques to ensure that only one wire is contacted for 
measurement.\cite{Wegrowe99,Jaccard00} We have developed techniques 
to remove nanowires from their templates\cite{Tanase01} and to make 
oxide-free electrical contacts to the ends of individual wires.\cite{Tanase02} 
In this work, we look at the effects of mechanically bending Ni nanowires. 
A nanowire with one bend has two segments whose magnetic easy axes point 
in different directions due to shape anisotropy. By varying the direction 
along which the wires are magnetized, a domain wall can be trapped at the bend. 
Electrical current passing through the nanowire flows through this trapped domain 
wall, allowing observation of effects due to the wall on the transport properties 
of the wire.

Previous work on domain wall magnetoresistance in nanostructures includes 
experimental studies on metal whiskers,\cite{Taylor68} template-grown 
nanowires,\cite{Radulescu00} GMR thin films,\cite{Gregg96} ferromagnetic thin 
films with stripe domains,\cite{Rudiger99} ferromagnetic trilayers,\cite{Prieto03} 
step-edge wires,\cite{Hong98} and lithographically defined structures,
\cite{Taniyama99,Rudiger98,Xu00} as well as theoretical models that incorporate 
spin-flip scattering,\cite{Levy97} spin accumulation,\cite{Simanek01} weak 
localization,\cite{Tatara97} and semi-classical scattering.\cite{vanGorkom99} 
The resistance contribution of domain walls  has been measured to be either 
positive\cite{Radulescu00,Gregg96,Rudiger99,Prieto03,Xu00} or 
negative\cite{Hong98,Taniyama99,Rudiger98} in various systems. The majority of 
these transport measurements were performed at low temperatures. In this work, 
howerver, we present results that show a decrease in resistance associated with 
the presence of a domain wall at room temperature, using mechanically bent 
nickel nanowires.

\section{Fabrication and Measurement}

\begin{figure}
\includegraphics[scale=4]{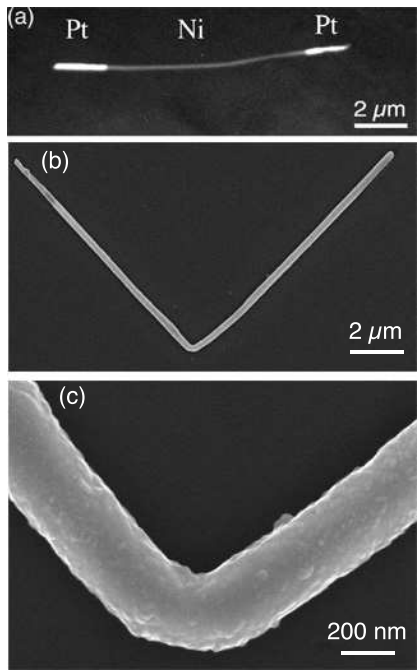}
\caption{\label{fig:SEM}SEM micrographs of PtNiPt nanowires. 
(a) Energy-resolved image of 3-segment nanowire showing Pt endcaps and 
Ni central segment. (b) Nanowire with $90^\circ$ bend at the center. 
(c) Detail of bend region for the wire shown in panel (b).}
\end{figure}

Metallic nanowires were grown via electrochemical deposition into the pores 
of nanoporous alumina templates.\cite{Possin70,Madsen85,Whitney93,Tanase01} 
The templates used were $60\:\mu\mathrm{m}$ thick with a nominal pore 
diameter of 100 nm (Anodisk, Whatman, Inc.), with a 500~nm thick sputtered 
copper film serving as a working electrode. Platinum was deposited from a 
solution of 7.3 g/L $\mathrm{(NH_4)_2PtCl_6}$ and 58.3 g/L $\mathrm{Na_2HPO_4}$, 
buffered to pH 8.3 at a potential of -0.6 V relative to a Ag/AgCl reference 
electrode. Nickel was deposited from a 
solution of 20 g/L $\mathrm{NiCl_2\cdot 6H_2O}$, 515 g/L 
$\mathrm{Ni(H_2NSO_3)_2\cdot 4H_2O}$, and 20 g/L $\mathrm{H_3BO_3}$, 
buffered to pH 3.4 at a potential of -1.0 V relative to a Ag/AgCl reference.  
Due to branching effects inside the pores, the actual diameters of the pores and 
hence of the nanowires, are significantly larger than the nominal diameter. 
Scanning electron microscopy (SEM) measurements on the nanowires showed 
diameters of $350\pm40$~nm. The nanowires were grown to lengths ranging 
from 20 to 40 microns. The majority of the wire length was nickel, with 
2~$\mu\mathrm{m}$ of platinum at each end of the wire (see Fig~\ref{fig:SEM}(a)). 
These Pt endcaps provide a clean low-resistance electrical interface between the 
nickel segment and the contacts.\cite{Tanase02}

\begin{figure}
\includegraphics[scale=4]{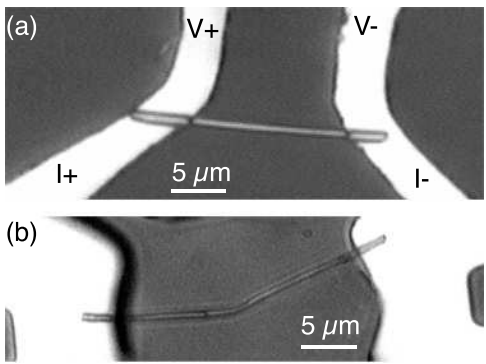}
\caption{\label{fig:patterned}Optical micrographs of PtNiPt nanowires 
with lithographic electrical contacts. (a) $20\:\mu\mathrm{m}$ long 
straight nanowire. (b) $35\:\mu\mathrm{m}$ long nanowire with $25^\circ$ bend.}
\end{figure}

After fabrication, the alumina template was dissolved in a $50\:^\circ$C 
KOH bath and the wires were resuspended in isopropanol. This suspension 
was then centrifuged for several minutes, inducing sharp bends at a range 
of angles into the nanowires. A SEM image of a nanowire with a $90^\circ$ 
bend is shown in Fig.~\ref{fig:SEM}(b), with a detail of the bend region shown 
in Fig.~\ref{fig:SEM}(c). This bending typically occurred near the center of the 
wire length. After bending, the wires were spun out onto a glass substrate, and 
an optical microscope equipped for projection photolithography\cite{Palmer73} 
was used to pattern evaporated Cr/Au electrical contacts on top of the platinum endcaps 
of the nanowires. Nanowires were selected  based on the angle and sharpness 
of the bend, and the straightness of the two segments. The contacts were patterned 
in a pseudo 4-probe geometry, as shown in Fig.~\ref{fig:patterned}.

Current-voltage (IV) measurements were performed on a series of straight 
PtNiPt nanowires of various lengths to determine the resistivity of the 
segments and the contact resistance. The IV curves were linear up to 
current densities of $j=5\times10^8\:\mathrm{A/cm^2}$ (I=0.5~mA), with 
Joule heating acting to increase the resistance at higher currents, up 
to a maximum breakdown current density of $j=10^{10}\:\mathrm{A/cm^2}$.\cite{Tanase02} 
The platinum segments showed a room-temperature resistivity of 
$\rho\approx 17\:\mu\Omega-\mathrm{cm}$  and the nickel had 
$\rho\approx 10\:\mu\Omega -\mathrm{cm}$.\cite{TanaseUnpub} 
Contact resistances between the electrodes and the nanowires 
were typically $1-2\:\Omega$. The measured resistances decreased 
monotonically as the wires were cooled to 5~K, indicating that the contacts 
were metallic in nature and all interfaces were clean. Similar measurements 
on pure Ni nanowires showed insulator-like thermal behavior, which suggests
the presence of surface oxide on the Ni and indicates
the importance of the Pt segments for the success of these measurements.

The magnetotransport measurements on PtNiPt nanowires were made with a 
$10\:\mu\mathrm{A}$ 100 Hz AC current source and a lockin amplifier. 
The measurements were made at room temperature, using an electromagnet 
equipped with a computer-controlled motorized rotating sample stage. 
A 2-axis Hall effect sensor (Sentron AG) was used to implement a closed 
feedback loop on the rotation system, giving an angular positional accuracy 
varying from $0.2^\circ$ at 4 kOe to $1^\circ$ at 500 Oe. Magnetoresistance 
hysteresis loops were obtained by ramping the field at rates ranging from 2 
to 50 Oe/s and continuously measuring the resistance. The system also allowed 
for measurement of resistance at constant field while continuously changing 
orientation. In this mode, the stage was rotated at 0.5 deg/s.  The feedback 
loop allowed for accurate comparisons between $R(H)$ and $R(\theta)$ measurements. 
In all cases, the field was in the plane of the substrate, with the rotation axis 
perpendicular to this plane.

\section{Straight nanowires}
\begin{figure}
\includegraphics[scale=0.35]{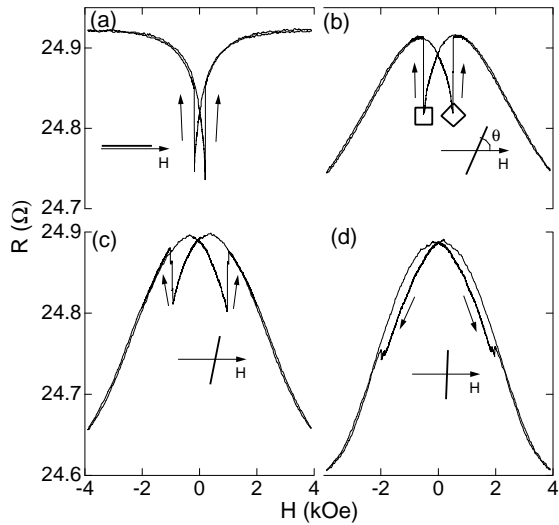}
\caption{\label{fig:straightMR}Magnetoresistance of a $20\:\mu\mathrm{m}$ 
long straight nanowire at different angles $\theta$ to the applied 
field. (a)--(d): R(H) at $\theta=0^\circ$, $70^\circ$, $82^\circ$, and $88^\circ$, 
respectively. The open square and diamond mark switching events at the same $(H,\theta)$ 
locations marked by the corresponding symbols on Fig.~\ref{fig:anglesweep_straight}(b). 
The arrows indicate direction of field sweep.}
\end{figure}

\begin{figure}
\includegraphics[scale=0.45]{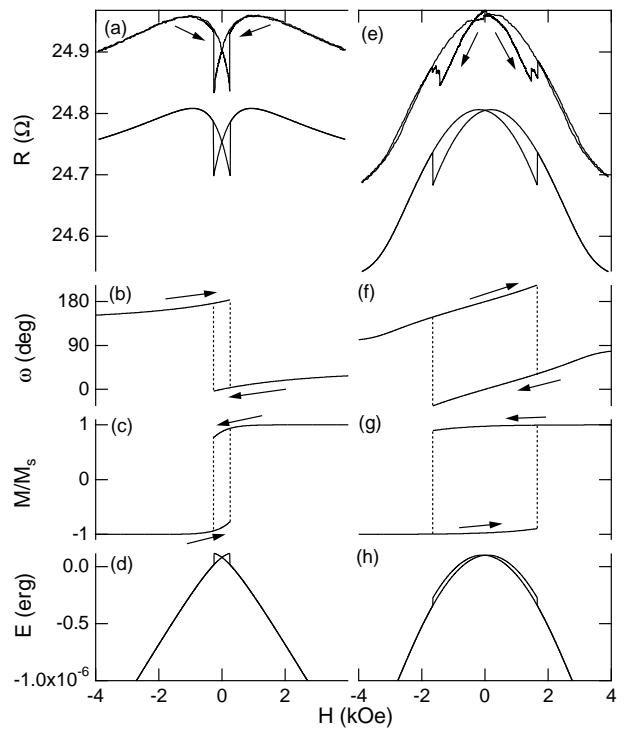}
\caption{\label{fig:straightE}Calculation of magnetoresistance and 
magnetostatic energy in straight nanowire using modified Stoner-Wohlfarth 
magnetization model. (a)--(d) Straight nanowire at $46^\circ$ to the 
field. (a) Measured and model magnetoresistances. Model is a best-fit 
curve to equations (\ref{eq:MH}) and (\ref{eq:straightMR}), 
with $H_0=1036$~Oe, $\Delta H=767$~Oe, 
$H_{\mathrm{sw}}=260$~Oe, $R_0=24.67\:\Omega$, and $\Delta R=0.306\:\Omega$. 
The model curve has been shifted downwards for clarity. Arrows show the direction 
of field sweep. (b) $\omega (H)$, the angle between the magnetization vector and the 
wire axis. Dashed lines show magnetization reversal at the curling mode switching 
field. (c) $M(H)$ for the model curve shown in (a).  (d) Magnetostatic energy 
calculated from (b) and (c). (e)--(h) Straight nanowire at $87^\circ$ to the 
field. $H_0=5067$~Oe, $\Delta H=2373$~Oe, $H_{\mathrm{sw}}=1650$~Oe.}
\end{figure}

The magnetoresistance of a 20~$\mu$m long straight nanowire is shown in 
Figs.~\ref{fig:straightMR}(a)-(d) for four different orientations to the field. 
As has been seen in previous, in-template, measurements on similar wires,
\cite{Wegrowe99,Jaccard00} there are two main features in the magnetoresistance. 
First, there is an overall non-hysteretic contribution due to the anisotropic 
magnetoresistance (AMR) effect in nickel.\cite{McGuire75} Our straight wires 
have a room temperature AMR of $\Delta R/R=(R(H=4\:\mathrm{kOe})-R(H=0))/R(H=0)=-1.5\%$ 
when the wire is perpendicular to the applied field. The decrease in the low-field 
resistance when the wire is parallel to the field (Fig.~\ref{fig:straightMR}(a)) 
indicates that there is some demagnetization at low fields. This is discussed in 
more detail below. In addition to the non-hysteretic magnetoresistance, there is 
an abrupt increase in the resistance at a well defined magnetic field, known as 
the switching field $H_{\mathrm{sw}}$.\cite{Wernsdorfer96} This hysteretic effect 
is due to the rapid reversal of the nanowire's magnetization.\cite{Wegrowe99} When 
the nanowire is nearly perpendicular to the field, we frequently observe a splitting 
of the transition into two or more unequal steps. This can be seen in 
Figs.~\ref{fig:straightMR}(c), (d) and \ref{fig:straightE}(e). We believe that the 
presence of the smaller transitions is due to slight bends in the nanowire, and a 
consequent shift in the switching field. When such split transitions are observed, 
we define the overall switching field as the location of the largest step in the 
resistance.

\begin{figure}
\includegraphics[scale=0.55]{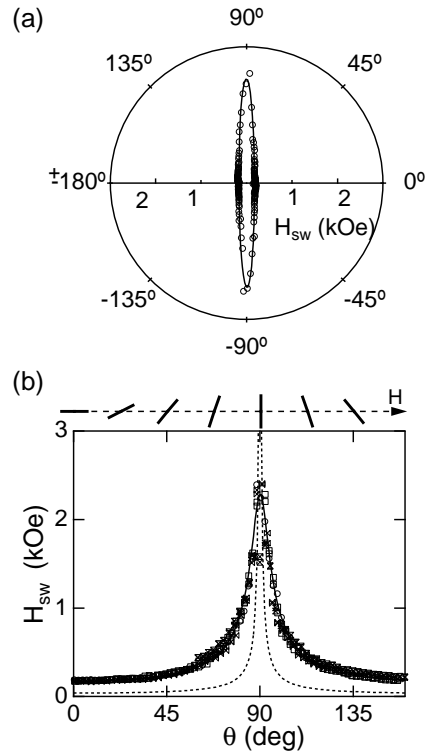}
\caption{\label{fig:straightHsw}(a) Polar plot of switching 
fields for a $20\:\mu\mathrm{m}$ straight wire. The external field is
applied in the $\theta=0$ direction; the nanowire points radially outward. 
Circles: Measured $H_{\mathrm{sw}}(\theta)$. Solid line: Curling small-nucleation-volume 
fit (see text). (b) Data and model from (a), replotted on linear scale, 
showing $|H_{\mathrm{sw}}|$. Switching fields from four nanowires with lengths 
ranging from 12 to 20~$\mu$m are plotted. Dashed line: Predicted $H_{\mathrm{sw}}$ 
for simultaneous reversal of the entire wire. The line drawings at the top of (b) 
show the direction of the wire orientation relative to the horizontal field.}
\end{figure}

As can be seen from Fig.~\ref{fig:straightMR}, the value of $H_{\mathrm{sw}}$ 
depends on the angle between the wire and the field. This is more evident in 
Fig.~\ref{fig:straightHsw}(a), which shows the angular dependence of the 
switching field. This figure can be viewed as a phase diagram for one 
particular nanowire. When a given $(H,\theta)$ location lies outside the 
elliptical region defined by the set of switching fields, the nanowire is in 
the reversible state, and the magnetization is single-valued, depending only
 on the values of $H$ and $\theta$, and not on the wire's prior history. On 
the other hand, when $(H,\theta)$ lies inside the set of switching points, 
the wire is in a hysteretic state, where the magnetization is multiply valued, and 
depends on the path followed. Fig.~\ref{fig:straightHsw}(b) shows the same switching 
field data as Fig.~\ref{fig:straightHsw}(a), plotted on a linear scale 
(showing $|H_{\mathrm{sw}}|$). In this panel, the reversible region lies above 
the plotted points, and the hysteretic region below. These measurements were 
repeated on a series of 350 nm diameter nanowires, with lengths ranging from 
12 to 20 $\mu$m. As can be seen from Fig.~\ref{fig:straightHsw}(b),  
$H_\mathrm{sw}(\theta)$ is independent of wire length in this range.

\begin{figure}
\includegraphics[scale=0.4]{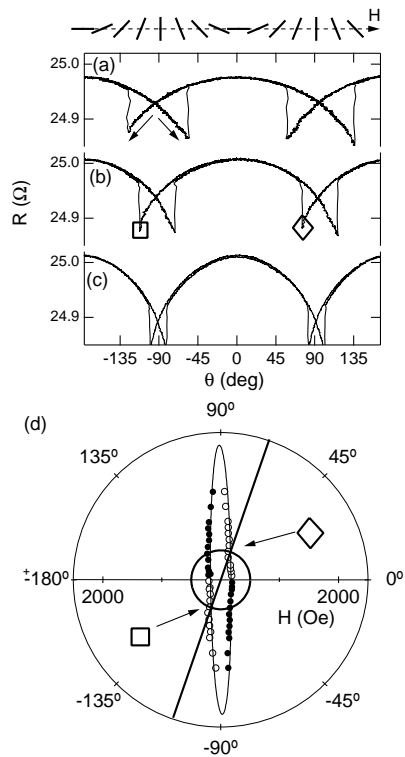}
\caption{\label{fig:anglesweep_straight}Resistance vs. wire orientation at 
fixed external field (a) H=300 Oe, (b) H=500 Oe, (c) H=1000 Oe for 
a $20\:\mu\mathrm{m}$ straight nanowire. The arrows show the direction of 
rotation. The line drawings above plot indicate wire orientation. The open 
diamond and square in panel (b) indicate switching events at the same $(H,\theta)$ 
locations marked by the corresponding symbols on Fig~\ref{fig:straightMR}(b). 
(d) Switching angle vs. applied field. The field points in the $\theta=0$ direction and 
the  nanowire points radially outward. The solid circles are from measurements with 
$\theta$ increasing, and open circles with $\theta$ decreasing. The solid curve is 
the curling-mode fit to the switching field data shown in Fig.~\ref{fig:straightHsw}. 
The straight line and circle show the trajectories in $(H,\theta)$ for the data in 
Fig.~\ref{fig:straightMR}(b) and panel (b) of this figure, respectively.}
\end{figure}

Additional measurements were performed where the field was kept fixed and the 
wire was continuously rotated through two complete revolutions, one 
counterclockwise ($\theta$ increasing) followed by one clockwise ($\theta$ decreasing). 
Three such $R(\theta)$ measurements on a straight nanowire are shown in 
Figs.~\ref{fig:anglesweep_straight}(a)--\ref{fig:anglesweep_straight}(c). The  
curvature seen in $R(\theta)$ is due to the magnetization of the wire rotating 
towards the direction of the field and hence away from the axis of the wire. The 
resistance follows $R\propto|\cos(\theta)|^2$, as would be expected from the AMR of a 
cylinder. The rotation sweeps also show sharp switching events that are very similar 
to those seen in the field sweeps discussed previously. These measurements also show 
hysteretic behavior. This can be understood by tracking the direction of the 
magnetization in the wire as it is rotated. Initially ($\theta=180^\circ$), the 
wire is parallel to the field, along with the magnetization. As the wire is rotated, 
the strong shape anisotropy tends to make the direction of {\bf M} rotate away from the 
field, tracking the wire axis. This continues as the wire rotates past perpendicular, 
with the magnetization becoming increasingly anti-aligned to the field. At a certain 
orientation ($\theta_{\mathrm{sw}}$), this anti-aligned magnetization becomes 
energetically unstable, and the magnetization undergoes a rapid reversal in direction 
to become aligned with the field. This produces the sharp increases in the resistance 
seen in Fig.~\ref{fig:anglesweep_straight}. When the direction of rotation is reversed, 
the same behavior is observed. However, since the reversal occurs after the nanowire 
rotates past perpendicular to the field, the magnetization, and hence the resistance, is 
hysteretic.

As with the $R(H)$ measurements discussed above, the $R(\theta)$ data can be divided 
into two regions, where the resistance is either reversible or hysteretic. The 
boundary between these two regions can be determined by plotting $\theta_{\mathrm{sw}}$ 
as a function of the applied field, as shown in Fig.~\ref{fig:anglesweep_straight}(d). 
The smooth curve in Fig.~\ref{fig:anglesweep_straight}(d) is the fit to 
the $H_{\mathrm{sw}}(\theta)$ data (discussed below) shown in Fig.~\ref{fig:straightHsw}. 
$H_{\mathrm{sw}}(\theta)$ and $\theta_{\mathrm{sw}}(H)$ closely overlap, indicating that 
the location of the phase boundary does not change depending on which variable is being 
swept. This is further illustrated by comparing the data shown in 
Fig.~\ref{fig:straightMR}(b) (R vs. H at $\theta=70^\circ$) with the data 
in Fig.~\ref{fig:anglesweep_straight}(b) (R vs. $\theta$ at H=500 Oe). The open 
diamond on both of these panels marks a switching event at $\theta=70^\circ$, H=500 Oe. 
Similarly, the open square marks a switching transition at $\theta=70^\circ$, H=-500 Oe, 
or equivalently, $\theta=-110^\circ$, H=500 Oe.

To determine the reversal mode of these nanowires, we examine the shape of 
$H_\mathrm{sw}(\theta)$. As shown in Fig.~\ref{fig:straightHsw}, 
$H_\mathrm{sw}$ is peaked near $\theta=\pm90^\circ$ (wire perpendicular 
to field) and has a minimum at $\theta=0,180^\circ$ (wire parallel to field). 
This angular dependence allows us to rule out Stoner-Wohlfarth, or coherent 
rotation, as the reversal mechanism, as that reversal mode has peaks in the 
switching field when the wire is both perpendicular and parallel to the 
field.\cite{Stoner48} This suggests that the reversal mechanism for the wire 
is incoherent rotation, or curling,\cite{Brown57} consistent with expectations 
based on the diameter of the nanowire,\cite{Aharoni00} as well as previous 
magnetic force microscopy studies on Ni nanowires of comparable size.\cite{Lederman95}

For an ellipsoid reversing its magnetization through the curling mode, it has 
been predicted\cite{Aharoni97} that the switching field follows
\begin{equation}
H_{\mathrm{sw}}=2\pi M_s \frac{(2D_z - \alpha) 
(2D_x-\alpha)}{\sqrt{(2D_z-\alpha)^2\sin^2\theta+(2D_x-\alpha)^2\cos^2\theta}}
\end{equation}
where $M_s$ is the saturation magnetization, $D_z$ and $D_x$ are the 
demagnetization factors along the major and minor axes of the ellipsoid, 
and $\alpha=k (r_0/r)^2$. Here, $r$ is the minor radius of the ellipsoid, 
$r_0$ is the exchange length of the ferromagnet, and $k$ is a geometrical 
factor dependent on the aspect ratio of the prolate spheroid. From the 
nanowire's length of $20\:\mu\mathrm{m}$ and radius $r=175\:\mathrm{nm}$, 
the demagnetization factors are $D_x=0.4998$ and $D_z=0.00042$.\cite{Morrish83} 
The saturation magnetization is assumed to be the bulk nickel value, 
$M_s=485\:\mathrm{Oe}$. The exchange length for nickel is known to be 
approximately 20 nm,\cite{Brown57,Lederman95} and $k$ for an extended 
cylinder is 1.079.\cite{Aharoni97} The predicted curve, shown as the 
dashed line in Fig.~\ref{fig:straightHsw}, clearly does not match the 
observed switching fields for our wires. However, it has been suggested that 
the magnetization reversal proceeds via an initial nucleation in a small volume 
of the wire, and subsequent propagation throughout the entire wire.\cite{Wegrowe99} 
This assumption does fit the observed data (solid line in 
Figs.~\ref{fig:straightHsw}(a), \ref{fig:straightHsw}(b), 
and \ref{fig:anglesweep_straight}(d)), with free parameters $D_z=0.0991$ 
and $\alpha=0.144$, corresponding to a nucleating region of aspect ratio 1.3:1 
and radius $r=100\:\mathrm{nm}$. This nucleation region occupies 0.5\% of the 
total wire volume. The size of this nucleation region is comparable to previously 
determined values in nanowires of smaller diameter.\cite{Jaccard00} As noted above, 
$H_\mathrm{sw}(\theta)$ for several of our straight nanowires have the same shape; 
this indicates that the size and shape of the initial nucleation volume does not 
vary signficantly between wires with different lengths.

To examine the energetics of the nanowire reversal process, we have developed a 
phenomenological model to describe the behavior of the magnetization. We begin by 
assuming that, except at the switching field, the direction of the magnetization is 
described by Stoner-Wolhfarth coherent rotation, 
\begin{equation}
\frac{H}{2\pi M_s}\sin(\theta-\omega)=(D_x-D_z)\sin(2\omega)
\end{equation}
where $\theta$ is the angle between the applied field and the wire axis, 
and $\omega$ is the angle between the magnetization vector and the wire 
axis.\cite{Aharoni00} Because the nanowire is initially being modeled as 
single domain, the demagnetization factors used in this calculation are 
derived from the geometry of the entire wire ($D_x=0.4998$, $D_z=0.00042$), 
and not from the nucleation volume determined from the curling mode fit. The 
resistance is then related to the magnetization by 
$R(H,\theta)=R_0 + \Delta R\cos^2\omega$. At $H=0$, the shape anisotropy 
forces the magnetization to lie along the axis of the wire ($\cos\omega=1$), 
indicating that the resistance should be maximized at zero field. However, 
the observed resistance data shown in Fig.~\ref{fig:straightMR} have maxima 
that are  at non-zero field. This implies that there is a demagnetization effect 
at low field that reduces the overall magnitude of {\bf M}. A possible mechanism 
for this effect is that the magnetization is rotating towards the crystalline easy 
axis of the wire. Since these nanowires are polycrystalline, there will be a 
distribution of these directions, resulting in a decrease in the net magnetization.

To model this behavior in the data, we assume that the S-W theory provides an 
accurate description of the direction of the magnetization, but that there is a 
spatially-uniform demagnetizing effect that reduces the magnitude. In the absence 
of direct data on the magnitude of the magnetization, we assume a simple analytic form,
\begin{equation}\label{eq:MH}
\frac{M(H)}{M_s}=\tanh\left(\frac{H\pm H_0}{\Delta H}\right)
\end{equation}
where the sign of $H_0$ depends on which branch of the hysteresis is being followed. 
The expression for the magnetoresistance is then modified to
\begin{equation}\label{eq:straightMR}
R(H,\Omega)=R_0 + \Delta R \left(\frac{M(H)}{M_s}\right)^2\cos^2\omega,
\end{equation}
and this expression is fit to the observed data using $H_0$, $\Delta H$, $R_0$, 
and $\Delta R$ as free parameters. We use the switching field calculated from the 
small nucleation volume curling mode fit described above, and not by the Stoner-Wohlfarth 
coherent rotation mode. The best-fit magnetization curves and calculated resistances 
for two orientations are shown in Fig.~\ref{fig:straightE}. The exact functional 
form of the demagnetization curve Eq. (\ref{eq:MH}) does not have a major effect 
on the quality of the model. A Langevin function ($L(x) = \frac{1}{x}-\coth(x)$) 
works as well as the $\tanh(x)$ that was used in Eq. (\ref{eq:MH}).

As can be seen from the figure, this approach produces calculated magnetoresistance 
curves that closely match the observed data, and hence this modeling procedure provides 
a good description of the magnetic response of the wire. The magnetization 
information obtained from the fit can then be used to calculate the magnetostatic 
energy of the nanowire. For an extended ellipsoid, the energy is given by
\begin{equation}\label{eq:Estraight}
E=-\frac{1}{2}V{\bf M}\cdot{\bf H} + \frac{1}{2}V4\pi(D_x-D_z)M_z^2
\end{equation}
where $V$ is the volume of the wire and $M_z$ is the component of the 
magnetization along the wire axis.\cite{Aharoni00} The field dependence of 
this energy is shown for two orientations of a straight wire in 
Fig.~\ref{fig:straightE}(c),(f). The magnetic reversal corresponds to a 
transition from a high-energy to a low-energy state, with a typical change 
in energy of $1\times10^{-7}$~erg, compared to a total magnetostatic energy 
on the order of $1\times10^{-6}$~erg.

\section{Bent nanowires}

\begin{figure}
\includegraphics[scale=0.4]{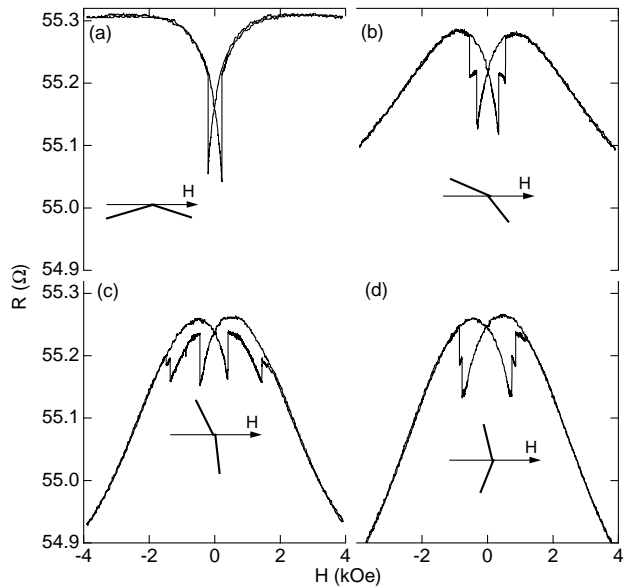}
\caption{\label{fig:bent30MR}Magnetoresistance of a $35\:\mu\mathrm{m}$ 
long $25^\circ$ bent nanowire at different angles to the applied field. (a) 
Both segments at $12^\circ$ to H ($\theta=180^\circ$ on Fig~\ref{fig:bent30Hsw}). 
(b) Segments at $41^\circ$ and $66^\circ$ to H ($\theta=126^\circ$). (c) Segments 
at $57^\circ$ and $82^\circ$ to H ($\theta=111^\circ$). (d) Segments at $77^\circ$ 
and $78^\circ$ to H ($\theta = 90^\circ$).}
\end{figure}

\begin{figure}
\includegraphics[scale=0.4]{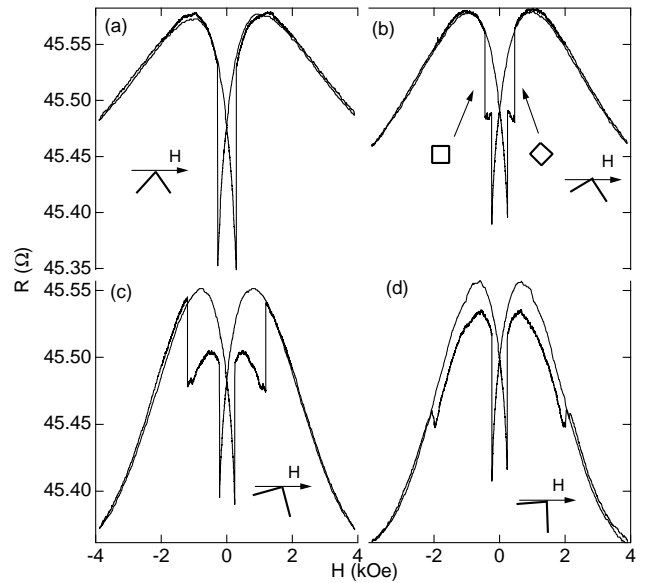}
\caption{\label{fig:bent90MR}Magnetoresistance R(H) of a $35\:\mu\mathrm{m}$ 
long $90^\circ$ bent nanowire at different angles to the applied field. (a) Both 
segments at $45^\circ$ to H ($\theta=180^\circ$ on Fig~\ref{fig:bent90Hsw}). (b) 
Segments at $20^\circ$ and $70^\circ$ to H ($\theta=155^\circ$). Open diamond and 
square mark switching events that occur at the same location as corresponding points 
marked on Fig.~\ref{fig:anglesweep90}(b). (c) Segments at $7^\circ$ and $83^\circ$ to 
H ($\theta=132^\circ$). (d) Segments at $2^\circ$ and $88^\circ$ to H ($\theta=137^\circ$).}
\end{figure}

\begin{figure}
\includegraphics[scale=0.75]{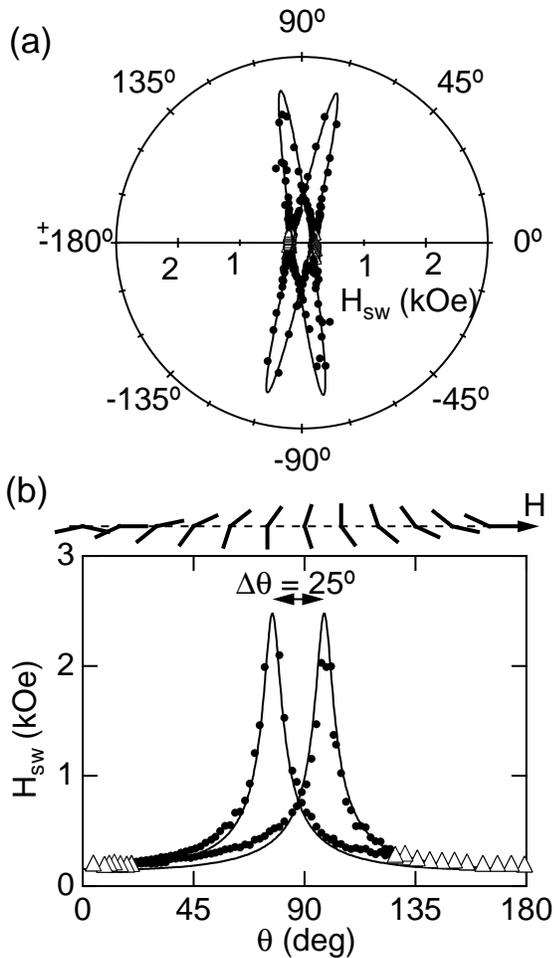}
\caption{\label{fig:bent30Hsw}Switching field versus wire orientation 
for a $25^\circ$ bent nanowire. The dots indicate the measured switching 
fields of the two segments; the open triangles show where only a single 
transition is observed. The solid lines are the small-nucleation-volume 
fit from Fig.~\ref{fig:straightHsw}, shifted in angle to account for the 
different orientations of the individual segments.}
\end{figure}

\begin{figure}
\includegraphics[scale=0.75]{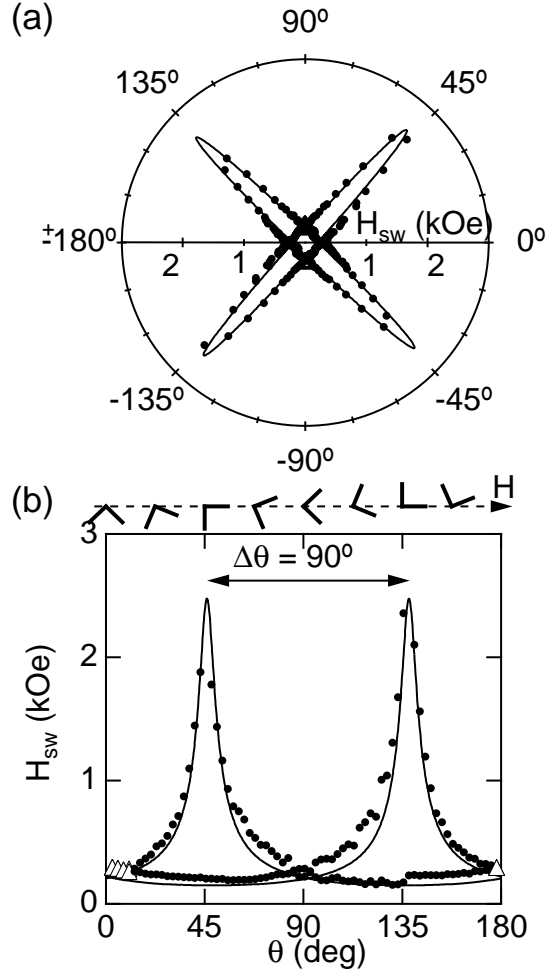}
\caption{\label{fig:bent90Hsw}Switching field versus wire orientation 
for a $90^\circ$ bent nanowire. The dots indicate the measured switching 
fields of the two segments; the open triangles show where only a single 
transition is observed. The solid lines are the small-nucleation-volume 
fit from Fig.~\ref{fig:straightHsw}, shifted in angle to account for the 
different orientations of the individual segments.}
\end{figure}

The magnetoresistance behavior of bent wires is signficantly different from 
that of the straight wires, as illustrated in Fig.~\ref{fig:bent30MR} for 
a 40~$\mu\mathrm{m}$ wire with a $25^\circ$ bend, and in Fig.~\ref{fig:bent90MR} 
for a wire of the same length with a $90^\circ$ bend. The non-hysteretic AMR 
component of the resistance is never flat, as it is for straight wires when 
$\theta=0$. In addition, most orientations show jumps in the resistance at 
two distinct fields, indicating the presence of two magnetization reversals. 
The switching fields vs. wire orientation for the $25^\circ$ and $90^\circ$ 
bent wires are shown in Fig.~\ref{fig:bent30Hsw} and Fig.~\ref{fig:bent90Hsw}. 
For each wire, there are two peaks in $H_{\mathrm{sw}}(\theta)$, each occuring 
when one of the segments is perpendicular to the field. As indicated in panel (b) 
of each figure, the separation between the peaks is equal to the bend angle for 
each wire. From this, the basic similarity of the individual peaks, and the comparable 
height of the resistance steps to that observed in the straight wires, we infer that 
the two resistance steps observed in $R(H)$ correspond to magnetization reversals in 
the individual segments, with the higher field feature coming from the segment that 
is oriented at the larger angle with respect to {\bf H}. It should be noted, however, 
that not all orientations show two switching transitions. For each bent wire, there 
is a range of orientations (marked by open triangles on Fig.~\ref{fig:bent30Hsw} 
and Fig.~\ref{fig:bent90Hsw}) where only a single transition is observed. The 
implications of this are discussed below.

As in the case of the straight nanowires, Fig.~\ref{fig:bent30Hsw} and 
Fig.~\ref{fig:bent90Hsw} represent phase diagrams for the bent nanowires. 
In this case, the shape of the phase boundaries is more complicated, because 
of the presence of two segments at differing orientations. The overall nature 
of the diagram, however, is the same, with a reversible region that lies outside 
the curve, and a hysteretic region inside the curve.

\begin{figure}
\includegraphics[scale=0.55]{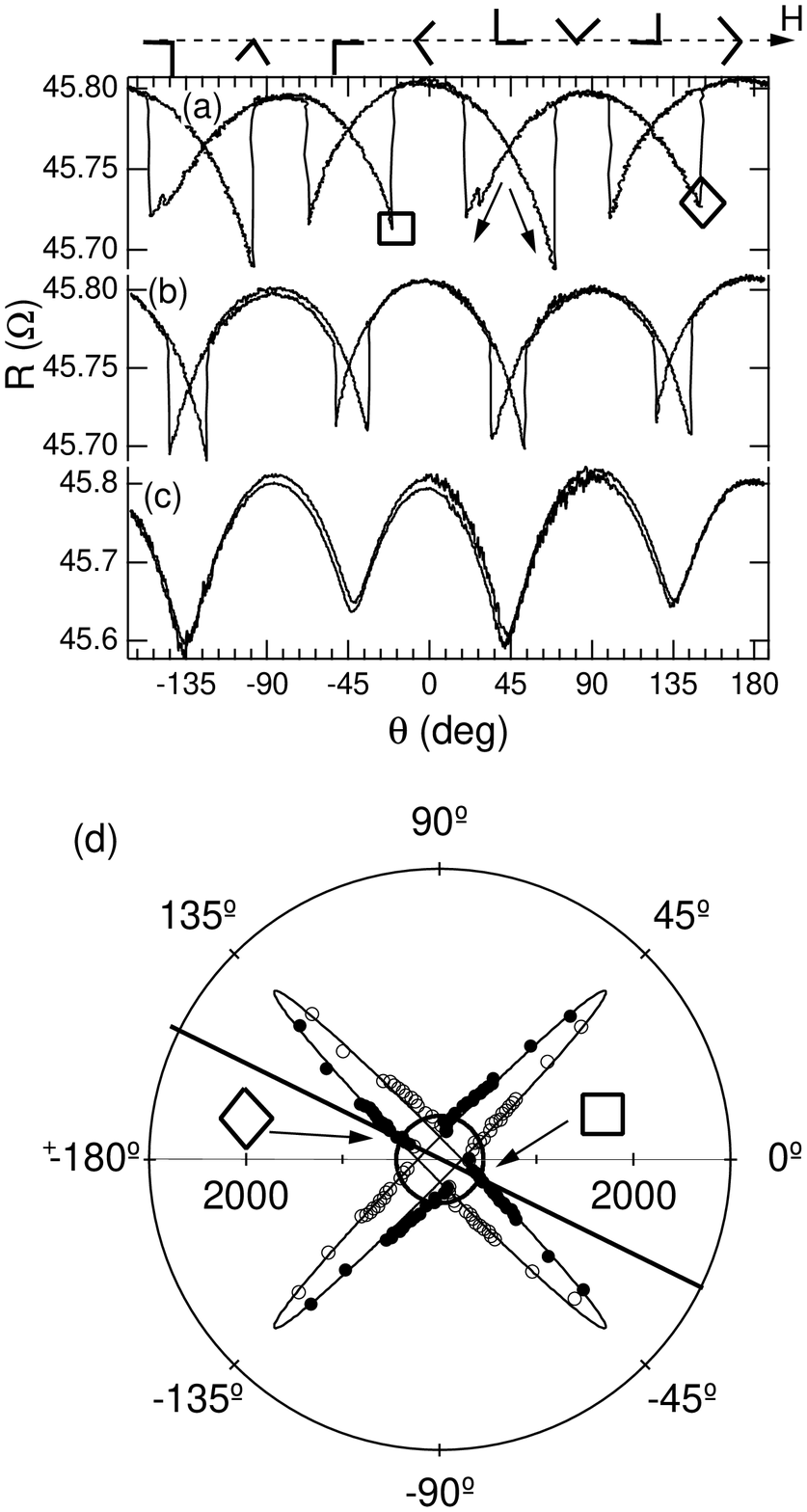}
\caption{\label{fig:anglesweep90}Resistance vs. wire orientation 
for a $40\:\mu\mathrm{m}$ nanowire with a $90^\circ$ bend at fixed 
external field (a) H=450 Oe. (b) H=1000 Oe. (c) H=3900 Oe. The arrows indicate the 
direction of rotation. (d) Switching angle vs. applied field. The  filled and open 
circles are switching transitions seen in counter-clockwise ($\theta$ increasing) 
and clockwise ($\theta$ decreasing) rotations, respectively. The solid line and 
circle represent trajectories followed in $(H,\theta)$ by Fig.~\ref{fig:bent90MR}(b) 
and in panel (a) of this figure, respectively. The open diamond and square indicate 
the position of switching events that occur on both trajectories.}
\end{figure}

Figure~\ref{fig:anglesweep90}(a)--(c) shows $R(\theta)$ measurements at a range of 
fields for a nanowire with a $90^\circ$ bend.  As in the straight wires, switching 
transitions in the bent wires are observed at the same set of $(H,\theta)$ locations 
in both the $R(H)$ and the $R(\theta)$ measurements. This is illustrated in 
Fig.~\ref{fig:anglesweep90}(d), where the smooth curves shown in 
Fig.~\ref{fig:bent90Hsw} are superimposed over the measured $\theta_{\mathrm{sw}}(H)$ 
from the $R(\theta)$ measurements. The two data sets match well. This is seen in more 
detail by comparing a field sweep at $\theta=155^\circ$ (Fig.~\ref{fig:bent90MR}(b)) 
to an angle sweep at H=450 Oe (Fig.~\ref{fig:anglesweep90}(b)). The trajectories in 
$(H,\theta)$ that these two data sets describe are marked in 
Fig.~\ref{fig:anglesweep90}(d) by a straight line and circle, respectively. 
The switching events on these two scans occur in the same location, indicated on 
the plots by an open diamond and square.

As a starting point, the angular dependence of the switching fields in the bent 
nanowires can be modeled by assuming that the two segments of the wire switch 
independently. If this assumption is accurate, then the curling-mode fit discussed 
above for the straight wire should also apply to bent wires when suitably offset in 
angle to account for the different relative orientation of the two segments to the 
field. As seen in Fig.~\ref{fig:bent30Hsw}(b) and Fig.~\ref{fig:bent90Hsw}(b), 
these curves qualitatively match the observed behavior of the bent wires, indicating 
that independent-segment curling is a reasonable first approximation for describing 
the switching behavior.\cite{Tanase03} There are, however, systematic deviations 
between the data and the model, indicating that the first-order approximation of 
the bent wire as two independent straight segments is incomplete. To understand 
these deviations, it is necessary to examine interaction effects between the segments.

\subsection{Effects of domain configuration on $H_{\mathrm{sw}}$}

\begin{figure}
\includegraphics[scale=2]{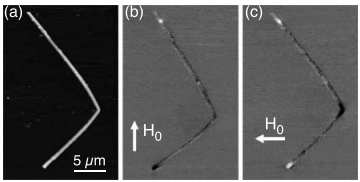}
\caption{\label{fig:MFM}(a) Atomic force microscopy image of a bent Ni nanowire. 
(b), (c) Magnetic force microscopy images of the same nanowire. In each panel, 
the wire was magnetized in the direction indicated by the arrow and subsequently 
imaged in zero field.}
\end{figure}

Before discussing the effects of intersegment interactions, it is first necessary 
to examine how the domain configuration in a bent nanowire varies with angular 
orientation. This can be examined using magnetic force microscopy (MFM) on bent 
nanowires. MFM images were acquired with a Nanoscope III Multimode AFM/MFM 
(Digital Instruments). An example is shown in Fig.~\ref{fig:MFM}. 
Fig.~\ref{fig:MFM}(a) shows an atomic force microscope topographic image. In 
Fig.~\ref{fig:MFM}(b), the nanowire was initially magnetized in a strong vertical 
field, and subsequently imaged by MFM at zero field. The MFM picture shows a positive 
pole at one end of the wire and a negative pole at the other end, with no pole at the 
bend. The absence of a pole in the body of the wire indicates that the magnetization 
is continuous, with no domain wall present. In Fig.~\ref{fig:MFM}(c), the same nanowire 
was remagnetized in a strong horizontal field and a zero-field image was again obtained. 
Here, the poles at the wire ends are both positive, and there is a strong negative pole 
at the bend. The pole at the bend is a signature of a the presence of a domain wall 
at the center of the wire. This central pole was measured to have a width 
of $1\:\mu\mathrm{m}$, significantly larger than the typical 100~nm domain 
wall width in nickel.\cite{Gregg96} We believe that this increase in the size 
of the domain wall is associated with the extended curvature at the bend region.

\begin{figure}
\includegraphics[scale=0.65,clip=true,bb=18 100 500 582]{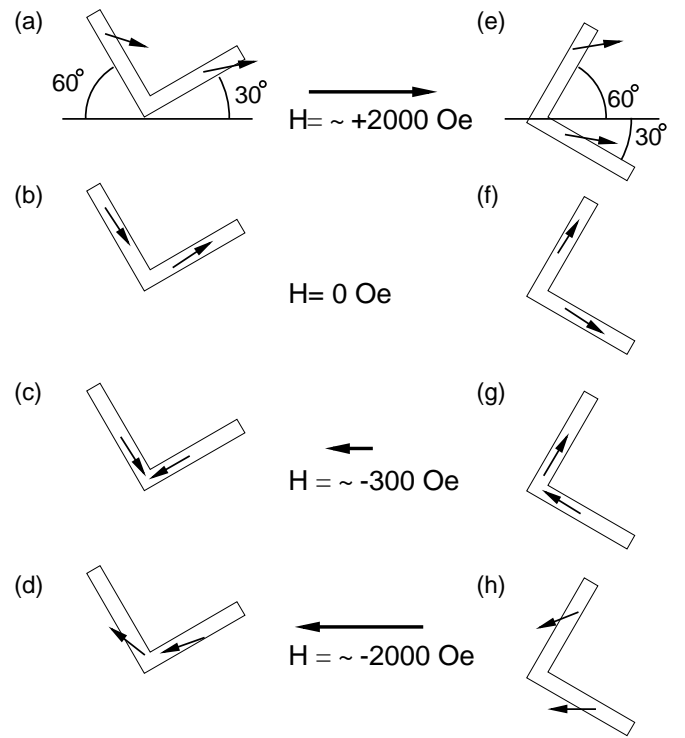}
\caption{\label{fig:orientation}Schematic of the magnetic configuration of a 
$90^\circ$ bent nanowire at two orientations to the field; the angles between 
the segments and the field are the same in both cases. (a)--(d) Nanowire 
in ``wall'' orientation (see text). (e)--(h) Nanowire in ``no-wall'' orientation. 
The applied field is given between each row. Note the presence of a domain wall 
in case c, but not in g.}
\end{figure}

The orientation dependence of the domain structure can be understood by considering 
the schematic of the magnetization structure shown in Fig.~\ref{fig:orientation}. 
At an initial large positive field (a,e), the applied field dominates over the shape 
anisotropy and the segments' magnetizations rotate off the segment axis and towards 
the applied field. When the field is reduced to zero, the magnetizations relax back 
towards the segment axes (b,f). The MFM images were taken at this point, with 
Fig.~\ref{fig:MFM}(b) corresponding to the schematic magnetization in 
Fig.~\ref{fig:orientation}(b), and similarly for Fig.~\ref{fig:MFM}(c) and 
Fig.~\ref{fig:orientation}(f). At some negative field, the segment that is nearly 
parallel to the field undergoes a switching transition, reversing its magnetization 
direction (c,g). At this point, the wire configuration shown in (c) has a $90^\circ$ 
domain wall at the bend; we therefore term this the ``wall'' orientation. On the other 
hand, the magnetization in (g) does not have a domain wall, but instead has a 
continuous rotation of the magnetization direction between the two segments. This 
orientation is thus designated the ``no-wall'' state. At larger negative field, 
the magnetization of the near-perpendicular segment reverses, and the magnetization 
for both segments again rotates off-axis (d,h) as in (a,e).

\begin{figure}
\includegraphics[scale=0.6]{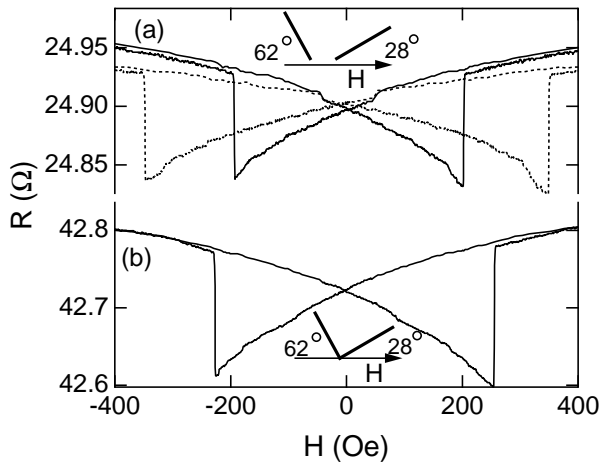}
\caption{\label{fig:locking} (a) R(H) of a straight nanowire
at $28^\circ$ (solid) and $62^\circ$ (dashed) to the field. (b) R(H) of
a $90^\circ$ bent wire, with segments at $28^\circ$ and $62^\circ$ to the field.}
\end{figure}

The most visible deviation from the independent segment picture occurs 
near $\theta=0^\circ$ and $\theta=180^\circ$, where the two segments are at similar 
angles to the field. If the two segments of the bent wire were not coupled, they would 
switch simultaneously if and only if their angles to the field were the same. For all 
other orientations, the segments would have different switching fields. In the actual 
bent wires, however, a single simultaneous transition is seen even when the angles are 
not the same. This is shown in Fig.~\ref{fig:locking}, which compares the  MR of a 
straight wire at $28^\circ$ and $62^\circ$ to the field to the resistance of a 
bent wire whose segments are at comparable orientations. The straight wire switches 
at different fields in the two orientations, whereas the bent wire's response is 
dominated by a single large jump in the resistance, indicating that both segments 
switch simultaneously. This ``locking'' of the switching of the segments occurs over 
a wide range of orientations, indicated by the solid triangles in 
Fig.~\ref{fig:bent30Hsw} and Fig.~\ref{fig:bent90Hsw}.

The reversal locking occurs in regions where the wire orientation is similar to 
the left column in Fig~\ref{fig:orientation}. This region is found at 
$\theta<78^\circ$ and $\theta>103^\circ$ in Fig.~\ref{fig:bent30Hsw} for the 
$25^\circ$ bent wire and $\theta<45^\circ$ and $\theta>135^\circ$ in 
Fig.~\ref{fig:bent90Hsw} for the $90^\circ$ wire. The locking occurs 
because the energy cost of nucleating a domain wall to enter the 
intermediate state (c) makes it  energetically favorable to suppress the separate 
transitions and reverse the entire wire simultaneously. The locked transitions 
are shown in Figs.~\ref{fig:bent30Hsw} and \ref{fig:bent90Hsw} as open triangles. 
An estimate for the energy cost of domain nucleation can be determined by considering 
the shift in the switching fields and the additional magnetostatic energy associated 
with that shift. For the orientations shown in Fig.~\ref{fig:locking}, this shift is 
approximately 35 Oe. This corresponds to an increase of energy of $1\times10^{-8}$~erg, 
approximately 1\% of the total energy of the nanowire.

When the two segments of a bent wire are at sufficiently different angles to 
the field, the wire reverses its magnetization in two separate steps. There is 
still an energy cost for switching into the anti-aligned intermediate state, 
which appears as an increase in the switching field of the segment closer to 
parallel to the field. This can be seen most clearly in Fig.~\ref{fig:bent90Hsw} 
for $\theta<45^\circ$ and $\theta>135^\circ$, where the measured switching 
fields (circles) are larger than what was expected based on the straight wire 
results (smooth curve). At higher field, the reversal of the segment closer to 
perpendicular to the field places the wire back into an aligned magnetization 
state. This is an energetically favorable transition, and in this regime, the 
straight wire data matches the behavior of the bent wire.

The opposite behavior is observed when the wire is in an orientation similar 
to the right column of Fig.~\ref{fig:orientation}. For the $90^\circ$ bent
wire, this occurs for $45^\circ<\theta<135^\circ$ in Fig.~\ref{fig:bent90Hsw}. 
In this orientation, the low-field reversal (of the segment more nearly parallel 
to the field) places the wire in a single domain state. Since there is no additional
energy barrier due to the creation of a domain wall, the behavior of the 
bent-wire switching field is accurately predicted by the straight wires. 
On the other hand, the subsequent reversal of the more nearly perpendicular 
segment places the wire in an anti-aligned magnetization state. The cost of 
nucleating a wall results here in an increase in the switching field in this 
regime. This is seen in Fig.~\ref{fig:bent90Hsw} as a shoulder in the peaks 
for $45^\circ<\theta<135^\circ$ compared to the prediction from the straight 
wire data.

It should be noted that these shifts in the switching fields for the bent 
nanowires cannot be ascribed to differences in the lengths of the two segments. 
The switching field data for the series of straight nanowires shown in 
Fig.~\ref{fig:straightHsw}(b) indicate that there is very little variation 
in $H_\mathrm{sw}(\theta)$ in the range of lengths studied. This range of 
lengths encloses the lengths of the segments of the bent nanowires studied. 
This implies that the changes in $H_\mathrm{sw}$ in the bent wires are not 
simply due to differences in the segment lengths.

\subsection{Domain wall resistance} 

\begin{figure}
\includegraphics[scale=0.55]{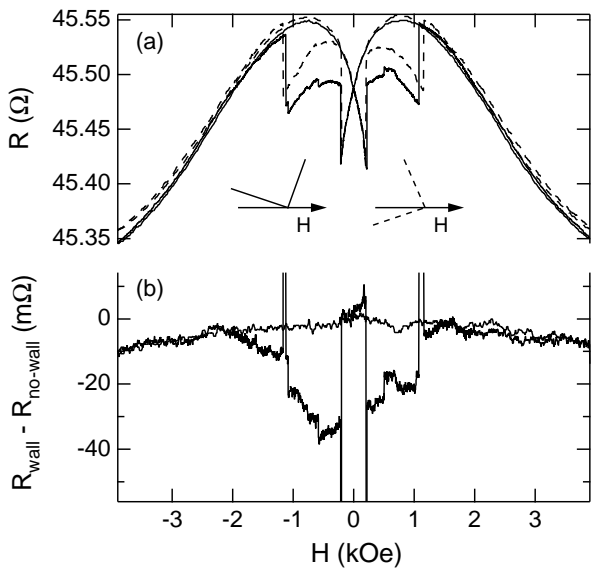}
\caption{\label{fig:wallR}Resistance versus field for a $90^\circ$ wire with 
segments at $8^\circ$ and $82^\circ$ to the field. (a) Solid line is the wire 
in the ``wall'' orientation (see text); dashed line is wire in the ``no-wall'' 
orientation. (b) Difference in resistance between the two wire orientations.}
\end{figure}

In addition to shifting the switching fields, the  domain configuration 
effects can be observed directly in the magnetoresistance data. This can 
be seen by comparing the MR curves for a bent wire in two orientations, 
corresponding to those in Fig.~\ref{fig:orientation}, with the same angles 
between the segments and the applied field. The magnetoresistance for one 
such pair of orientations is shown in Fig.~\ref{fig:wallR}(a). The solid curve, 
corresponding to the orientation in the left column of 
Fig.~\ref{fig:orientation} (``wall'' orientation), has a domain wall in 
the region where only one segment has reversed. The dashed curve, corresponding 
to the other (``no-wall'') orientation, does not have a domain wall in this region. 
Since the angles between the segments and the field are the same, we would expect 
the magnetoresistances to be the same, with some differences in the switching 
fields due to the energetic effects discussed above. Instead, there is a 
significant difference in the intermediate region, where only one of the 
segments has reversed. This is shown in more detail in Fig.~\ref{fig:wallR}(b), 
which plots the differences between the two curves. The solid curve 
(``wall'' orientation) has a lower resistance in the intermediate region, 
suggesting that there is a decrease in the resistance associated with the 
presence of a domain wall. This decrease was seen in measurements on 
multiple $90^\circ$ bent wires, with magnitudes in the 
range of $20-30\:\mathrm{m}\Omega$, corresponding to a fractional change in the 
resistance of $\Delta R/R=-5\times10^{-4}$. We note that the bend angle needs 
to be near $90^\circ$ to allow a comparison of this sort to be made. This result 
is somewhat analogous to what was observed by Taniyama \emph{et. al} in 
lithographically-defined zigzag structures at lower temperatures.\cite{Taniyama99}

There are multiple theoretical models that include the possibility of 
domain walls with negative contributions to the resistance. Tatara and 
Fukuyama\cite{Tatara97} have proposed a mechanism based on the suppression of 
weak localization. Using the dimensions of our nanowire in their theory gives 
a predicted wall resistance of $\delta R/R=-4\times10^{-4}$, comparable to the 
measured value. However, it is unlikely that weak localization effects play a 
significant role at room temperature, indicating that an alternate explanation 
is required. van Gorkom \emph{et al.}\cite{vanGorkom99} have proposed a 
semi-classical mechanism based around differing scattering relaxation times for the 
majority and minority spin channels. Depending on the ratio of these times, the wall 
resistance can be either negative or positive, with values in the 
range $-0.05<\delta R/R<0.1$. This range encompasses the observed values 
in our nanowires, but in the absence of detailed information on the nature 
and density of the impurities in the nickel, we are unable to make a more 
precise comparison between the theory and our data.

Another effect that can provide negative wall resistance is the AMR contribution 
from the spins inside the wall itself.\cite{Hong98,Taniyama99} Based on the MFM 
results shown in Fig.~\ref{fig:MFM}, the domain wall width in the bent wires is 
on the order of $1\:\mu\mathrm{m}$. Given the straight wire satuartion 
magnetoresistance of $\Delta R/R=-1.5\%$, this corresponds to a decrease 
in the resistance of $8\:\mathrm{m}\Omega$, approximately one third of the 
decrease actually observed. Thus, while it appears that AMR plays a significant 
role in the resistance shift, it seems likely that there are other mechanisms 
contributing to this effect and that further investigation is necessary to 
determine its cause.

\section{Conclusions}
In summary, we examined the magnetotransport properties of straight and bent 
ferromagnetic nanowires. It was determined that the switching behavior of the 
straight nanowires is consistent with the curling-mode reversal of a small 
volume, followed by propagation throughout the wire bulk. The bent wires 
showed qualitatively similar behavior, with modifications from the intersegment 
interactions. For both the straight and bent nanowires, the location of 
the switching events in $(H,\theta)$ were found to be independent of which 
variable was being swept. The magnetic properties of the bent wires were 
dependent on the domain configuration at the bend; the energetic cost of 
nucleating a domain wall acts to increasd the observed switching field compared 
to a straight wire at equivalent orientations. There is also a change in the 
resistance associated with the domain configuration; the resistance is lower 
when a domain wall is present. Further investigation is needed to determine 
the mechanism of this reduction, including examination of wires of differing 
diameters and possible studies of the dynamics of the reversal. Thus it appears 
that magnetic nanowires are a fruitful system for future studies of reversal 
dynamics and the effects of domain walls in ferromagnets.

\section{Acknowledgments}
We thank M. Stiles and P. Searson for illuminating discussions.
This work was supported by NSF Grant No. DMR--0080031.

\end{document}